\documentclass[aip,jcp,reprint,amsmath,amssymb,floatfix,citeautoscript,nofootinbib]{revtex4-2}
\usepackage{cancel}
\usepackage{amsmath}
\usepackage{physics}
\usepackage{graphicx}
\usepackage{amssymb}
\usepackage{amsthm}
\usepackage[T1]{fontenc}
\usepackage[version=4]{mhchem}
\usepackage{bm}
\usepackage{float}
\usepackage{dcolumn}
\newcolumntype{x}[1]{D{.}{.}{#1}}

\usepackage{ragged2e}
\usepackage{lipsum}
\usepackage{newtxtext,newtxmath}
\allowdisplaybreaks

\usepackage{color}
\usepackage[usenames,dvipsnames]{xcolor}
\definecolor{myblue}{rgb}{0,0,1}
\usepackage[breaklinks=true,colorlinks=true,linkcolor=myblue,urlcolor=myblue,citecolor=myblue]{hyperref}

\let\vr\undefined
\newcommand{\vr}{{\bm{r}}}

\let\Oe\undefined
\newcommand{\Oe}{O$_\mathrm{e}$}
\newcommand{\Cc}{C$_\mathrm{c}$}
\newcommand{\Ce}{C$_\mathrm{e}$}

\usepackage{tabularx}

\begin{document}

\title{Reaction dynamics of lithium-mediated electrolyte decomposition using machine learning potentials}

\author{Sohang Kundu}
\email{sohangkundu@gmail.com}
\affiliation{Department of Chemistry, Columbia University, New York, NY 10027, USA}

\author{Diana Chamaki}
\affiliation{Department of Chemistry, Columbia University, New York, NY 10027, USA}

\author{Hong-Zhou Ye}
\affiliation{Department of Chemistry and Biochemistry, University of Maryland, College Park, MD, 20742, USA}
\affiliation{Institute for Physical Science and Technology, University of Maryland, College Park, MD, 20742, USA}

\author{Garvit Agarwal}
\affiliation{Schr{\"o}dinger, Inc., New York, NY 10036, USA}

\author{Timothy C. Berkelbach}
\email{t.berkelbach@columbia.edu}
\affiliation{Department of Chemistry, Columbia University, New York, NY 10027, USA}
\affiliation{Initiative for Computational Catalysis, Flatiron Institute, New York, NY 10010, USA}

\begin{abstract}

We study the ring-opening decomposition of ethylene carbonate in the presence
of a single lithium atom and on the surface of lithium metal. Combining
accurate electronic structure theory, enhanced sampling, and machine learning,
we fine-tune the MACE-MP0 foundation model and apply the resulting machine
learning potentials to obtain statistically converged free energy profiles and
reaction rates.  We confirm that the level of electronic structure theory is
important, and inaccurate density functionals can overestimate the reaction
rate by up to nine orders of magnitude. We also find that harmonic transition
state theory underestimates reaction rates by about one order of magnitude. For
the surface reaction, we find and characterize a new, ultrafast decomposition
pathway wherein the carbonyl is deeply inserted into the lithium surface and
bent by about 70$^\circ$. This reaction, which occurs in a few tens of
picoseconds, generates a ring-opened intermediate that is a precursor for CO or
\ce{CO2} formation; by contrast, an alternative pathway that yields \ce{CO3^2-}
and ethylene is found to be non-competitive, occurring on a timescale of tens of
nanoseconds.

\end{abstract}

\maketitle

\section {Introduction}

The solid electrolyte interphase (SEI) is a passivation layer formed in
lithium-ion batteries due to decomposition reactions occurring
at the interface between the anode and the electrolyte. The SEI plays an important
functional role because it allows ion transport while
preventing further electrolyte decomposition\cite{Peled2017,Wang2018}. However,
uncontrolled SEI growth is detrimental as it consumes active materials, leads
to dendrite formation, and compromises the Coulombic efficiency of the
battery\cite{Lin2017,Weng2019}. Designing materials to control SEI composition
and growth for optimal battery performance is currently an active area of
research\cite{Li2017,Pathak2020}, which would benefit from a detailed
understanding of the mechanism for SEI formation. 

Over the past twenty years, many computational
studies\cite{Wang2001,Leung2011,Young2018,CamachoForero2015,Ebadi2016,Debnath2023,Li2025}
have investigated the reductive decomposition of organic carbonates, such as
ethylene carbonate (EC), in the presence of lithium to understand the initial
steps of SEI formation. However, many of these studies have focused on
characterizing only a few reactant and product geometries along with the
transition state or minimum energy path connecting
them~\cite{Wang2001,Ebadi2016,Debnath2023,Agarwal2025}; with these inputs, harmonic
transition state theory (TST) provides a simple estimate of the reaction rate.  At
realistic operating temperatures, reaction dynamics can be significantly more
complicated, which can alter the reaction mechanisms and the predicted rates of
reaction. Therefore, some studies have performed ab initio molecular dynamics
(MD) simulations, but the computational costs have limited them to relatively
low levels of electronic structure theory, small system sizes, and/or a few
short trajectories~\cite{Leung2010,Leung2011,Young2018}. 

Here we perform several fully atomistic, finite-temperature MD 
studies of the decomposition of EC in the presence of lithium, leveraging
developments in machine-learning potentials (MLPs) to improve the electronic structure
description, the configurational sampling, and the total simulation time.
Specifically, we focus on the initial ring-opening reactions for two systems.
The first system is a Li-EC molecular complex, which is a model for reductive
decomposition in the presence of a lithium ion near the anode surface.  The
second system is a single EC molecule on a periodic (001) surface of lithium metal.
For convenience, we refer to these as the `molecular' and `surface' reactions,
respectively.
Our use of MLPs enables us to efficiently compute exact free energies and rates
via fully atomistic simulations on accurate, reactive potential energy
surfaces.

Our objectives in this work are threefold. First, we aim to investigate how finite-temperature
statistical and dynamical effects influence the mechanism of EC decomposition.
Second, we aim to develop and validate a protocol for the training and application 
of MLPs on ab initio data for decomposition reactions and surface chemistry.
Third, we aim to evaluate the accuracy of commonly used computational and theoretical
approximations when computing reaction energy profiles and rates.  
Our findings provide valuable insights into the chemical dynamics of electrolyte
decomposition reactions and offer guidance for the accurate modeling of
reactions at interfaces. 

\section{Results}

\begin{figure}[t]
   \includegraphics[scale=1]{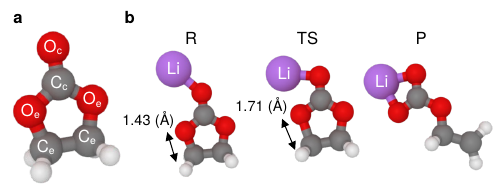}
   \caption{(a) Molecular structure of ethylene carbonate (EC), with carbon and oxygen labels used throughout the text.
(b) Reactant (R), transition state (TS), and product (P) geometries for the initial ring-opening
reaction of LiEC.
}
   \label{fig:molecular_structure}
\end{figure}

The EC molecule contains a five-membered ring with four C–O bonds and one C–C
bond. Under reducing conditions, the various C–O bonds become labile, leading
to different decomposition reactions.  Following previous
works~\cite{Leung2011,Young2018,Agarwal2025}, we denote the two chemically
distinct C-O bonds as {\Ce}-{\Oe} (ethylene carbon to ether oxygen) and
{\Cc}-{\Oe} (carbonyl carbon to ether oxygen), noting that there are two
equivalent bonds of each type,
as shown in Fig.~\ref{fig:molecular_structure}a. 

In what follows, we focus exclusively on the initial ring-opening step of all
reactions. Subsequent decomposition or diffusion of the product species is
assumed not to influence the rate of the ring-opening reaction. To ensure that
these pathways do not contribute to the calculated free energies, a soft wall
potential was applied during our simulations, restricting
the sum of the two equivalent {\Ce}-{\Oe} bonds and the sum of the two
equivalent {\Cc}-{\Oe} bonds to remain below 5~\AA\ at
all times. Additionally, a softer wall potential was applied to prevent
desorption of EC or its decomposition products from the Li atom or the Li
surface. 

\subsection{Molecular reaction}

In lithium-ion batteries, EC molecules coordinate with \ce{Li+} ions in
solution.  After a one-electron transfer to reduce \ce{Li+EC}, one of the
{\Ce}-{\Oe} bonds is cleaved, to generate a ring-opened radical, as shown in
Fig.~\ref{fig:molecular_structure}b.  Two subsequent pathways have been
characterized: the radicals can dimerize to form lithium ethylene dicarbonate
and ethylene, or a second electron transfer can trigger cleavage of the other
{\Ce}-{\Oe} bond, yielding \ce{LiCO3^2-} and ethylene. Because these latter two
pathways require several EC molecules or additional electrons, we do not study
them here.

This reaction network was characterized in seminal work by Balbuena et
al.~\cite{Wang2001} using density functional theory (DFT) with the B3PW91
functional. Recently, the accuracy of various DFT functionals was assessed for
the energetics of this reaction~\cite{Debnath2023} by comparing against higher levels of theory,
including quantum Monte Carlo and coupled-cluster theory with single, double, and perturbative triple
excitations [CCSD(T)]. In that work, dispersion-corrected
range-separated hybrid functionals were found to perform especially well, with
errors of 2--3~kcal/mol. In particular, barrier heights are accurately
predicted, whereas semilocal functionals and global hybrids underpredict
barrier heights by 5--10~kcal/mol, which would imply reaction rates that are
far too large.  

We study this first ring-opening reaction at three levels of electronic
structure theory: PBE-D3~\cite{Perdew1996,Grimme2010},
$\omega$B97X-D3~\cite{Lin2012}, and CCSD(T). All ab initio calculations are
performed using ORCA~\cite{Neese2011,Neese2020}, and the CCSD(T) results are
obtained with the domain-based local pair natural orbital
approximation~\cite{Riplinger2013,Riplinger2013a}.  To facilitate downstream
calculations, including geometry optimization, normal-mode analysis, and MD
simulations, we train three MLPs to these levels of theory.  Our MLPs are
trained by fine-tuning of the MACE~\cite{Batatia2022,Batatia2025} foundation
model (MP0)~\cite{Batatia2024}, which is a strategy that has been shown to
lower the amount of training data when compared to training from
scratch~\cite{Kaur2025}.  Importantly, we generate our training data by
umbrella sampling along the {\Ce}-{\Oe} bonds, ensuring that the training data
includes a diversity of configurations, including those drawn from the
reactant, transition state, and product ensembles.  
For information on dataset generation and MLP fine-tuning, see the Methods
section and the Supplementary Information. 

\begin{table}
\caption{\label{tab:molecular}
Reaction barrier heights (in kcal/mol), reaction rates (in s$^{-1}$), and
recrossing correction $\kappa$ for the ring opening reaction of LiEC. Values
indicated by ``harm'' are calculated in the harmonic approximation with
vibrational normal-mode frequencies. All results are presented for MLPs
trained to the indicated level of theory.
   }
\begin{ruledtabular}
\begin{tabular}{lcccccc}
Theory          & $\Delta E^\ddagger$ &  $\Delta F_\mathrm{harm}^\ddagger$ & $k^\mathrm{TST}_\mathrm{harm}$ & $\Delta F^\ddagger$ & $k^\mathrm{TST}$ & $\kappa$ \\ 
\hline
PBE-D3          &  4.8 &  5.1 & $1.2\times 10^9$ &  4.2 & $2.1\times 10^{10}$ & 0.49 \\  
$\omega$B97X-D3 & 18.3 & 17.5 & $1.1\times 10^0$ & 16.0 & $8.3\times 10^1$    & 0.94 \\ 
CCSD(T)         & 16.6 & 16.1 & $1.2\times 10^1$ & 15.4 & $1.6\times 10^2$    & 0.96 \\ 
\end{tabular}
\end{ruledtabular}
\end{table}

In Tab.~\ref{tab:molecular}, we present the energy barrier heights $\Delta
E^\ddagger$ predicted by each MLP, defined as the electronic energy difference
between the optimized transition state and reactant geometries.  As expected,
PBE-D3 predicts a barrier height that is much lower than that of
$\omega$B97X-D3 or CCSD(T), which agree with each other to about 1.5~kcal/mol.
We calculated the $T=300$~K free energy barriers $\Delta F^\ddagger_\mathrm{harm}$ 
(obtained using harmonic vibrational frequencies) and the associated
harmonic TST reaction rates $k^\mathrm{TST}_\mathrm{harm}$,
both of which are given
in Tab.~\ref{tab:molecular}. The predicted rates vary over nine orders of magnitude,
primarily due to the differences in the electronic energy barriers and not the vibrational
contributions.

\begin{figure*}[t]
   \includegraphics[scale=1]{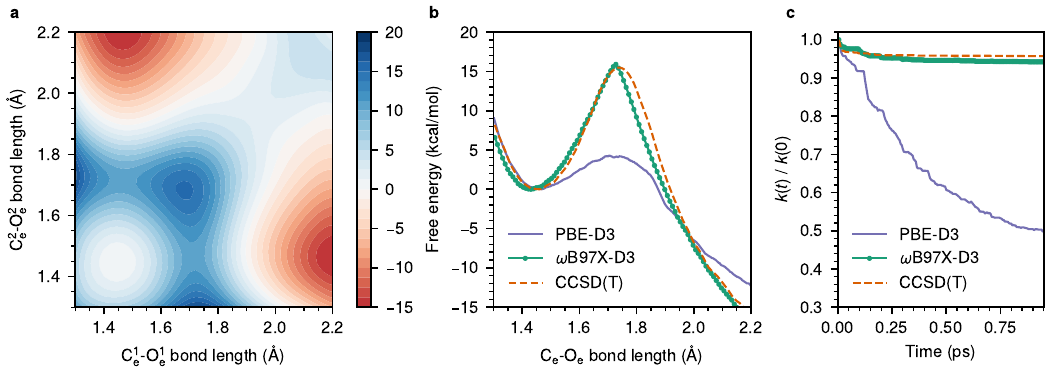}
   \caption{(a) Free energy surface of a single EC molecule as a function of its two chemically equivalent
{\Ce}-{\Oe} bond lengths using the $\omega$B97X-D3 MLP. (b) Free energy profile along one {\Ce}-{\Oe} bond length
from each MLP. (c) Normalized flux-side correlation function for each MLP, quantifying the recrossing
corrections to transition state theory.}
   \label{fig:molecular_data}
\end{figure*}

Next, we assess the accuracy of the harmonic free energy barriers and reaction
rates by comparing them to exact values obtained by MD.
We performed a 1.5~ns well-tempered metadynamics~\cite{Barducci2008}
simulation using the $\omega$B97X-D3 model. 
We biased both of the chemically equivalent {\Ce}-{\Oe} bond distances,
generating the two-dimensional free energy surface (FES) 
shown in Fig.~\ref{fig:molecular_data}a.
The 2D FES suggests that the two bond-breaking processes are largely independent, 
and either one can be studied in isolation.
In what follows, we perform umbrella sampling using 
one of the {\Ce}-{\Oe} bond lengths as a reaction coordinate (RC),
whose validity we have confirmed by calculating the average committor and its
distribution along the RC, as shown in the SI. 
The free energy profile for all three MLPs is shown in Fig.~\ref{fig:molecular_data}b.
Compared to the harmonic free energy barriers, the exact free energy barriers are
all lower by about 1~kcal/mol,
and the exact free energy barrier is lower than the
0~K TS energy barrier by about 1--2~kcal/mol; these findings hold for all MLPs.
The slight reduction in barrier heights is consistent with the idea that the 
TS, with its partially broken bonds,
is floppier.
Interestingly, although the $\omega$B97X-D3 and CCSD(T) models predict similar
energy and free energy barrier heights, the free energy profiles are quite different:
the $\omega$B97X-D3 free energy barrier is much more narrow than that of CCSD(T),
which might be expected to yield different reaction dynamics.

Next, we calculate the reaction rate using the TST approximation and using the
exact, long-time limit of the flux-side correlation function $k(t)$. 
This TST rate is the zero-time limit and thus a purely statistical
approximation to the exact rate, and it improves upon the harmonic TST rate by
including all configurational anharmonicity.  The recrossing factor that
corrects the TST rate is defined by $\kappa = k_\mathrm{exact}/k_\mathrm{TST}$.
In practice, we evaluate these rates using importance sampling via the
Bennett-Chandler method~\cite{Bennett1977,Chandler1978}.

In Fig.~\ref{fig:molecular_data}c, we show the normalized flux-side correlation function
$k(t)/k(0)$, and in Tab.~\ref{tab:molecular}, we present the TST rate and the
recrossing correction factor $\kappa$ for each model.
The anharmonic TST rates are 10--100 times larger than the harmonic TST rates,
primarily because the anharmonic free energy barriers are smaller than the harmonic ones.
The $\omega$B97X-D3 and CCSD(T) models show
$\kappa \approx 1$, indicating that the adiabatic approximation is
good and that dynamical recrossings are minimal. 
In contrast, the PBE-D3 level
shows a significantly lower $\kappa$, resulting in a reduction of the TST rate
by a factor of two. This stronger recrossing effect is attributed
to a lower free energy barrier and a flatter curvature at the FES maximum. 
Based on our most accurate models, we predict EC decomposition timescales of
$k^{-1} \sim 10$~$\mu$s, although an explicit solvent can be expected to modify
this prediction.  More specifically, we believe $\kappa$ is large because our
calculations are performed in the gas phase, and solvent friction would reduce
these values of $\kappa$.

\subsection{Surface reaction}

\begin{figure*}[t]
   \includegraphics[scale=1]{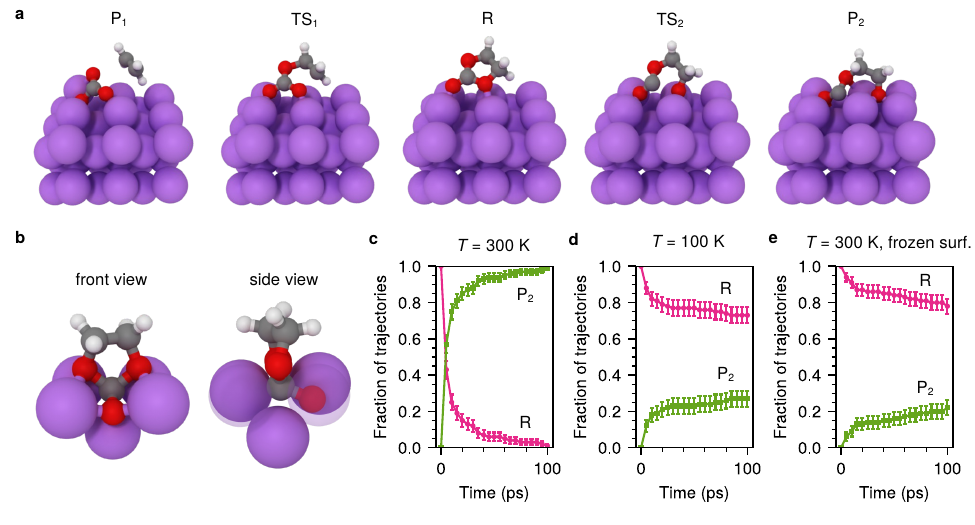}
   \caption{(a) Geometries of the adsorbed EC reactant (R) in the parallel motif,
the two transition states (TS1 and TS2), and the two products (P1 and P2),
associated with cleavage of the {\Ce}-{\Oe} bonds and the {\Cc}-{\Oe} bond, respectively.
(b) Geometry of the `bent' adsorption minimum, highlighting the deep carbonyl insertion, which
is representative of structures observed in MD simulations of {\Cc}-{\Oe} bond breaking.
(c), (d), (e) Fraction of trajectories that remain as reactant (R, pink) or break a {\Cc}-{\Oe}
bond towards product (P$_2$, green) for simulations performed at 300~K, 100~K, and 300~K with the
surface atoms frozen. Error bars show the standard deviation of the mean.
All simulations use an MLP trained to the PBE-D3 level of theory.
}
   \label{fig:surface}
\end{figure*}

We next study the reactivity of EC on the (001) surface of neutral, unbiased
lithium.  Unlike for molecular chemistry, the best level of electronic
structure theory for surface chemistry on metals in unclear and an active area
of research. For example, although nonlocal exchange is understood to improve
barrier heights for molecular reaction, it worsens the description of bulk
metals and their surfaces~\cite{Paier2007,Stroppa2008}. Moreover, CCSD(T), the
``gold standard'' of molecular quantum chemistry, is inapplicable to metals due
to an infrared divergence~\cite{Shepherd2013,Masios2023,Neufeld2023}.
Therefore, we limit our study of the surface reaction to the use of PBE-D3.
Remarkably, PBE, which was also used in early AIMD studies of EC
decomposition~\cite{Leung2010,Leung2011}, was recently found to be one of the
best performing functionals on a database of dissociative chemisorption barrier
heights on transition metal surfaces~\cite{Tchakoua2022}, with a mean absolute
error of only 2.4~kcal/mol.

In the so-called `parallel' motif, the EC molecule adsorbs to the surface via
the interaction of the carbonyl oxygen and the ether oxygen with two lithium
atoms on the surface, as shown as the reactant (R) in Fig.~\ref{fig:surface}a.
This parallel geometry has been found in previous studies to be more stable
than other local minima, such as the `top' and `bridge'
geometries~\cite{Ebadi2016}.  NEB calculations identify two decomposition
pathways associated with the cleavage of the {\Ce}-{\Oe} (as in the molecular
reaction) or the {\Cc}-{\Oe} bond~\cite{Agarwal2025}. The former pathway leads
directly to CO$_3^{2-}$ on the surface and C$_2$H$_4$ gas, with two electrons
transferred from the metal surface.  The latter pathway leads to a stable
ring-opened product, which could undergo further decomposition towards CO or
CO$_2$ byproducts.  The two transition states (TS$_1$ and TS$_2$) and products
(P$_1$ and P$_2$) are shown in Fig.~\ref{fig:surface}a.  Although early studies
of the molecular reaction~\cite{Wang2001,Wang2002} suggested that the first
pathway would dominate initial SEI formation, later work studying the surface
reaction~\cite{Leung2010,Leung2011} found evidence that the second pathway
dominates.  Like for the molecular reaction, we generated PBE-D3 fine-tuning
data for an MLP by ab initio MD with umbrella sampling---in this case, along
the two distinct bonds.  All following results are those of the MLP.

For the four-layer slab used in this work, the reaction barrier of the first
pathway (breaking {\Ce}-{\Oe}) is $\Delta E^\ddagger = 6.1$~kcal/mol,
and that of the second pathway (breaking {\Cc}-{\Oe}) is 
$\Delta E^\ddagger = 5.2$~kcal/mol.  Given these comparable barrier heights,
one may expect similar reaction rates.  However, MD simulations at 300~K
initiated from the reactant geometry found that all reactive trajectories broke
the {\Cc}-{\Oe} bond almost immediately (forming the product P$_2$ in
Fig.~\ref{fig:surface}a) and never the {\Ce}-{\Oe} bond.  Most trajectories
showed decomposition within only 10--20~ps, as shown in
Fig.~\ref{fig:surface}c.

To understand the origin of this rapid and selective decomposition, we
inspected the trajectories and found that many of the reactions proceeded
through a slightly different mechanism than that implied by the transition
state TS$_2$.  By using configurations from reactive trajectories as initial
guesses to additional NEB calculations, we identified several additional paths
with barriers as low as $\Delta E^\ddagger = 0.7$~kcal/mol.  A unifying feature
of these additional paths is a deeper insertion of the carbonyl into the
lithium surface.  In fact, by minimizing the energy of 800 configurations
randomly sampled from the reactive trajectories, we identified a new local
energy minimum in which the carbonyl inserts into a hollow site, with puckering
of the ring and significant bending of the carbonyl by about 70$^\circ$, as
shown in Fig.~\ref{fig:surface}b.  Although this `bent' minimum energy
structure was found with a MLP, we confirmed its stability with ab initio
PBE-D3 calculations, and the adsorption energy is about 
40~kcal/mol lower than that of the original adsorbed geometry.  To the best of
our knowledge, this is the first report of this bent adsorption motif,
complementing the top, bridge, and parallel motifs.  Although the reactive
trajectories do not strictly follow a two-step mechanism passing through this
minimum (the barrier from the new `bent' minimum to the product P$_2$ is about
4.2~kcal/mol), they follow a qualitative similar one-step pathway.  We
hypothesize that the deep insertion into the lithium surface followed by
bending of the carbonyl destabilizes the EC ring, triggering {\Cc}-{\Oe} bond
breaking.

The deep insertion occurring in these reactive trajectories requires
significant accommodation by the surface lithium atoms, whose fluctuations are
dictated by the temperature.  Indeed, MD simulations performed at 100~K show
reduced reactivity: although about 20\% of trajectories show fast
decomposition, most of the remainder are nonreactive even after 100~ps
(Fig.~\ref{fig:surface}d).  To further validate the importance of this
insertion mechanism, we repeated MD simulations at 300~K with all lithium atoms
kept frozen at their initial positions (Fig.~\ref{fig:surface}e).  As expected,
we find that the reactivity is significantly suppressed: after 100~ps, only
about 20\% of trajectories have exhibited {\Cc}-{\Oe} bond breaking.

We conclude that, on the surface of bare lithium, decomposition via cleavage of
the {\Cc}-{\Oe} bond to form a ring-opened structure occurs almost immediately
(within 10--20~ps) and outcompetes the alternative decomposition pathway. The
reaction occurs so quickly that the adsorbed EC molecule is not a metastable
reactant, and a reaction rate cannot be meaningfully defined. Although the fast
reactivity precludes a detailed study of the slower {\Ce}-{\Oe} bond-breaking
mechanism, we can study the latter by constraining the {\Cc}-{\Oe} bond length
to prevent reactivity. Under this constraint, the adsorbed geometry is
metastable and we can perform a reaction rate study analogous to the one we
performed for the molecular reaction.

\begin{figure}[t]
   \includegraphics[scale=1]{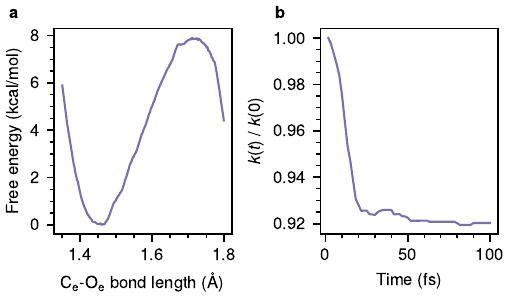}
   \caption{(a) Free energy surface and (b) normalized flux-side correlation
function for {\Ce}-{\Oe} bond breaking on the lithium surface shown in
Fig.~\ref{fig:surface}, with the {\Cc}-{\Oe} bond length constrained to
preclude reactivity. Results were obtained at 300~K with an MLP trained to the
PBE-D3 level of theory.
}
   \label{fig:CeOe}
\end{figure}

In Fig.~\ref{fig:CeOe}a, we show the calculated free energy profile for
{\Ce}-{\Oe} bond breaking (committor analysis confirmed the validity of the
bond length as a reaction coordinate, as shown in the SI).  The free energy
barrier is $\Delta F^\ddagger = 8.0$~kcal/mol, which is higher than the
electronic energy barrier obtained by NEB and also higher than the free energy
barrier of the molecular reaction, which breaks the same bond.  From anharmonic
TST, we calculate a reaction rate of 
$k_\mathrm{TST} = 4.7\times 10^7$~s$^{-1}$. In Fig.~\ref{fig:CeOe}b, we present
the normalized flux-side correlation function, which shows a recrossing
correction of $\kappa = 0.92$ and an exact reaction rate of 
$4.3\times 10^7$~s$^{-1}$. From the large free energy barrier and associated
20~ns reaction timescale, we conclude that the {\Ce}-{\Oe} bond breaking is not
remotely competitive with {\Cc}-{\Oe} bond breaking.

\section{Discussion}

To summarize, we have demonstrated how MLPs can be efficiently trained and
applied to chemical reactions, focusing on the decomposition of EC in the
presence of lithium. Such MLPs have allowed us to rigorously evaluate the
impact of the electronic structure theory and common approximations for
reaction rates, for both molecular (gas phase) and surface reactions. Within
the approximations taken in this work, we can estimate that the molecular
reaction occurs in about 10~$\mu$s, and that the surface
reaction occurs in about 20~ps and proceeds almost exclusively via
breaking of the {\Cc}-{\Oe} bond, leading towards CO or CO$_2$ products.
This latter pathway was assigned a mechanism based on surface fluctuations that
facilitate insertion and bending of the EC carbonyl, which may also promote
subsequent production of CO.

The present work is only a first step meant to systematically evaluate
improvements in the electronic structure theory and the sampling. Future work
must consider explicit solvent, the voltage of the anode, and the dynamics of
the electron transfer that triggers decomposition, which we expect to vary from
the early to late stages of SEI formation. Accounting for any one of these via
MLPs is not straightforward but is the subject of ongoing research.  We expect
rapid progress, perhaps via charge-aware
MLPs~\cite{Ko2021,Dajnowicz2022,Deng2023} and/or non-adiabatic MD with
MLPs~\cite{Chen2018,Dral2018,Westermayr2020}. The present workflow can also be
combined with hybrid Monte Carlo schemes that allow access to longer
timescales~\cite{Takenaka2014,SpotteSmith2022}.

\section{Methods}

\subsection{Ab initio calculations}

All ab initio molecular dynamics (AIMD) simulations were performed with Quantum
Espresso \cite{Giannozzi2009,Giannozzi2017}, using the PBE
functional\cite{Perdew1996} with D3 dispersion \cite{Grimme2010} and PAW
pseudopotentials \cite{Kresse1999}. A plane-wave basis was used with kinetic
energy cutoffs of 40~$\mathrm{E_h}$ (wavefunctions) and 160~$\mathrm{E_h}$
(density), and Gaussian smearing of $0.005~\mathrm{E_h}$ for finite-temperature
occupations. For the molecular reaction, a cubic cell of length $15$~\AA~ was
used to prevent image interactions. For the surface reactions, the Li surface
was modeled as a $3\times3\times2$ slab with four atomic layers in the
$z$ direction, plus $15$~\AA~of vacuum, yielding a
$10.13$~\AA$~\times~10.13$~\AA$~\times~21.76$~\AA~supercell. AIMD trajectories
used a $2\times2\times1$ $k$-point mesh and a 1~fs timestep, and biasing was performed 
with umbrella potentials using the Plugin for Molecular Dynamics
(PLUMED) \cite{Tribello2014}.

From the AIMD data, about $4500$ configurations were chosen for fine-tuning,
for which single-point evaluations were repeated at the targeted levels of
theory.  For the molecular reaction, we performed DFT calculations in the
def2-TZVP basis~\cite{Schaefer1994,Weigend2005} and DLPNO-CCSD(T) calculations
in the cc-pVTZ~\cite{Dunning1989} basis, both using ORCA~\cite{Neese2020}.  For
the surface reaction, we performed PBE-D3 calculations with a denser
$3\times3\times1$ $k$-point mesh using Quantum Espresso. 

\subsection{Machine learning potentials}

To train our MLPs, we iteratively grew the training datasets by using MD from
intermediate MLPs trained to each target level of theory.  For both the
molecular and surface reactions, we used about 4500 configurations with a 91:9
training:validation split, and tested the resulting MLPs on about 3000 unseen
configurations.  For the molecular reaction, the root-mean-square errors in the
energies are less than 6~meV/atom (1.5~kcal/mol total) for both DFT models and
about~8 meV/atom (2.0~kcal/mol total) for the CCSD(T) model, which was trained
without forces.  For the surface reaction, the root-mean-square error in the
energy is about 3 meV/atom.

\subsection{Free energies and reaction rates}

Within the harmonic approximation, free energies for all MLPs were calculated as
\begin{equation}
\Delta F_\mathrm{harm}^\ddagger = \Delta E^\ddagger 
    - k_\mathrm{B}T \ln Q_\mathrm{vib}^\mathrm{TS}/Q_\mathrm{vib}^\mathrm{R}
\end{equation}
where
\begin{equation}
Q_\mathrm{vib} 
    = \prod_i^{\omega_i^2 > 0} \left(1-e^{-\hbar\omega_i/k_\mathrm{B}T}\right)^{-1},
\end{equation}
and $\omega_i$ are normal-mode frequencies.
The harmonic TST reaction rate is
\begin{equation}
k^\mathrm{TST}_\mathrm{harm} 
    = \frac{k_\mathrm{B}T}{h} e^{-\Delta F_\mathrm{harm}^\ddagger/k_\mathrm{B}T}.
\end{equation}

The exact free energy profile along reaction coordinate $s$ was calculated as
\begin{equation}
F(s) = -k_\mathrm{B}T \ln \int d^{3N}r\ \delta[s(\vr)-s] e^{-V(\vr)/k_\mathrm{B}T}
\end{equation}
and the exact free energy barrier $\Delta F^\ddagger$ is the difference between the barrier
maximum and the reactant minimum (the dividing surface $s^*$ where the free
energy is maximized was found to coincide with that determined by committor analysis).

To calculate the exact reaction rate, we use a dividing
surface $s^*$ to separate the reaction coordinate $s$ into reactants and
products and calculate
\begin{equation}
k(t) = \frac{1}{\langle 1-h_\mathrm{P}\rangle} \langle \dot{s}\delta(s-s^*) h_\mathrm{P}(t) \rangle
\end{equation}
where $h_\mathrm{P} = \theta(s-s^*)$ is the product indicator function,
from which
$k_\mathrm{TST} = k(t\rightarrow 0^+)$ and
$k_\mathrm{exact} = \lim_{t\rightarrow t_\mathrm{p}} k(t)$
where $t_\mathrm{p}$ is a plateau time.  

\section{Data Availability}
The ab initio data used for fine-tuning and the result MLP will be made
available for public use at the time of publication.

\section*{Acknowledgements}
We thank Glen Hocky, Jinggang Lan, and David Limmer for helpful discussions.
This work was funded by the Columbia Center for Computational Electrochemistry.
We acknowledge computing resources from Columbia University’s Shared Research
Computing Facility project, which is supported by NIH Research Facility
Improvement Grant 1G20RR030893-01, and associated funds from the New York State
Empire State Development, Division of Science Technology and Innovation
(NYSTAR) Contract C090171, both awarded April 15, 2010.
The Flatiron Institute is a division of the Simons Foundation.

\end{document}